# Improving Computational Efficiency in DSMC Simulations of Vacuum Gas Dynamics with a Fixed Number of Particles per Cell

Moslem Sabouri[1], Ramin Zakeri[1], and Amin Ebrahimi[2,*]

[1] *Faculty of Mechanical Engineering, Shahrood University of Technology, P.O. Box 36199-95161, Shahrood, Iran*

[2] *Faculty of Mechanical Engineering, Delft University of Technology, Mekelweg 2, 2628CD Delft, The Netherlands*

* Corresponding author: A.Ebrahimi@tudelft.nl (A. Ebrahimi)

**Abstract**

The present study addresses the challenge of enhancing computational efficiency without compromising accuracy in numerical simulations of vacuum gas dynamics using the direct simulation Monte Carlo (DSMC) method. A technique termed "fixed particle per cell (FPPC)" was employed, which enforces a fixed number of simulator particles across all computational cells. The proposed technique eliminates the need for real-time adjustment of particle weights during simulation, reducing calculation time. Using the SPARTA solver, simulations of rarefied gas flow in a micromixer and rarefied supersonic airflow around a cylinder were conducted to validate the proposed technique. Results demonstrate that applying the FPPC technique effectively reduces computational costs while yielding results comparable to conventional DSMC implementations. Additionally, the application of local grid refinement coupled with the FPPC technique was investigated. The results show that integrating local grid refinement with the FPPC technique enables accurate prediction of flow behaviour in regions with significant gradients. These findings highlight the efficacy of the proposed technique in improving the accuracy and efficiency of numerical simulations of complex vacuum gas dynamics at a reduced computational cost.

**Keywords:** Rarefied gas flow; Direct simulation Monte Carlo (DSMC); Computational efficiency; Microchannel mixer; External supersonic flow.

# 1. Introduction

The direct simulation Monte Carlo (DSMC) method [1] is a computational technique for studying vacuum gas dynamics, particularly in scenarios in which the continuum assumption of fluid dynamics breaks down. This method is essential for understanding and predicting the behaviour of gases at low pressures and high Knudsen numbers, where interactions between gas molecules and solid surfaces play a significant role in the flow dynamics [2-5]. In DSMC simulations of gas flows, the gas is represented by a group of particles (often called simulator particles) that move through a computational domain. These particles represent a statistical number of real gas molecules, and their positions and velocities are tracked over time [6]. The equivalent number of molecules ($F_{num}$) indicates to the number of real gas molecules represented by each simulator particle. DSMC simulations progress in discrete time steps. During each time step, particles move ballistically (*i.e.* without collisions) until a collision event is encountered or they exit the computational domain [4]. DSMC accounts for the molecular nature of dilute gases by simulating binary collisions between particles. The probability of collision is determined based on the relative velocities and the collision cross-section, which depends on the specific gas being simulated [7]. When two particles collide, the model accounts for the transfer of momentum and energy between them based on conservation laws. Post-collision velocities are determined using stochastic methods, and the particles continue their motion accordingly. To facilitate efficient calculations, the computational domain is divided into smaller cells (*i.e.* computational grid). Collisions between particles are limited to those in the same cell or neighbouring cells. Finally, statistical sampling is performed to estimate macroscopic properties of the gas, such as density, velocity, temperature, and pressure. A more comprehensive description of the DSMC method can be located in other references [1-4]; thus it is not repeated in this manuscript.

Like any other numerical method, certain numerical parameters can affect the accuracy of the results obtained from DSMC simulations. Factors such as the time step size and the computational cell size are also important in this method. The effects of these parameters have been investigated extensively in previous studies (see for instance [8-10]). A critical numerical parameter specific to the particle-based methods, such as the DSMC method, is the number of simulator particles in each computational cell, which is often called the number of particles per cell (PPC). Considering the fluctuations of this parameter over time, the average value of PPC is critical to ensure the accuracy and reliability of DSMC predictions. Previous investigations have discussed the effect of PPC on the accuracy of calculations [10-16].

In a DSMC simulation, the number of particles may not be equal for all computational cells. In fact, factors such as variations in number density (*i.e.* number of molecules per unit volume) and differences in cell dimensions in different parts of the computational domain can lead to variations in the number of particles per cell. The number of particles per cell (PPC) is directly proportional to the number density and cell volume, and inversely proportional to the equivalent number of molecules ($F_{num}$). The difference in the number of particles in different cells can be significant in some problems. For instance, in the study of the flow passing through a conduit into a vacuum environment, there is a very sharp density drop from the conduit's inlet to its outlet. This can result in a much smaller number of PPC in the cells adjacent to the outlet compared to the cells close to the inlet. A similar issue arises when the computational grid resolution is increased in specific regions of the physical domain to capture intense local gradients (*e.g.* shock waves [17,18] or flow passages with sudden expansion or contractions [19,20]). This reduction in the number of PPC has two consequences. Firstly, the number of simulator particles in cells located in low density regions or refined cells may decrease below the required level to maintain the physical accuracy of the results. Secondly, a decrease in the number of simulator particles implies reduction in the statistical sample size, resulting in unphysical fluctuations in predicted macroscopic quantities. Although reducing the equivalent number of molecules ($F_{num}$) can increase the number of simulator particles in underpopulated cells, it also increases the number of molecules in the other cells (possibly beyond the necessary level), leading to an unnecessary increase in the computational cost.

Managing computational costs while maintaining accuracy is crucial in simulations involving a wide range of local Knudsen numbers or significant variations in local gradients. Various approaches have been explored, including CFD-DSMC hybrid methods [17,21-23] and all-molecular hybrid methods such as DSMC-Equilibrium Particle Simulation Method (EPSM) [24], DSMC-Low Diffusion (LD) [25], DSMC-Dynamic Collision Limiter [26], and DSMC-Fokker-Planck [27-29] hybrid methods. These methods address challenges arising from significant variations in local rarefaction, concentrating primarily on the increased computational cost in DSMC simulations with decreasing local Knudsen numbers. However, the present study focuses on additional difficulties in DSMC simulations, specifically those related to significant variations in the number of simulator particles per cell resulting from changes in cell size or local density. A potential resolution to these challenges involves maintaining a uniform distribution of particles across all computational cells, thereby providing enhanced control over both the physical and statistical accuracy, as well as the computational cost, of a DSMC simulation.

The concept of employing per-species weight factors has been previously suggested in the simulation of mixture flows containing trace species [30-36]. This methodology addresses numerical challenges caused by the low concentration of particles representing minor constituents in mixtures. However, this is not the concern of current study. The current study aims at regulating the total number of particles per cell, irrespective of the species type.

The concept of regulating the number of simulator particles in computational cells has been previously implemented in simulations using an axisymmetric formulation in DSMC [1]. In axisymmetric simulations, cell volumes are proportionate to their distance from the symmetry axis. Uniform cell divisions along the radial direction lead to increased particle accumulation in cells located farther from the axis compared to those near the axis. To address this, assigning different weights to particles in various cells controls the variations in particle numbers along the radial direction. This weighting is similar to defining the "equivalent number of molecules" parameter as a function of the radial position of the cell (*i.e.* distance from the symmetry axis). This technique has already been incorporated into the open-source software SPARTA [5,37], which also allows for weight assignment proportional to the cell volume. While this approach addresses variations in cell volume, it fails to achieve a uniform distribution of particles across the computational domain in every simulation.

Variable particle weighting has been used in particle-based simulations to manage the number of simulator particles, thereby reducing statistical errors in scenarios with severe density gradients or when adaptive grids are employed. Among pioneering works is the study by Kannenberg and Boyd [38], wherein a three-dimensional DSMC simulation was conducted to simulate thruster plume impingement on satellite solar panels. They employed axisymmetric formulations for the near thruster region and three-dimensional formulations for the far field, with the nearfield region providing inlet boundary conditions for the far field. Subdividing the far field into three regions, they applied weight factors proportional to 0.1, 0.4, and 1 of the reference weight, respectively, based on the region's distance from the inlet. This weighting strategy correlates with the inverse proportionality of cell dimensions to density, resulting in varying particle densities in each cell by controlling the number of particles per cell, thus leading to a non-uniform PPC distribution across the computational domain.

While this weighting strategy mitigated the severity of PPC variations across the computational domain, its goal was not to achieve a uniform PPC distribution.

Several studies have employed the "split and merge" strategy to regulate the number of particles, with the objective of maintaining the conservation of mass, momentum, and energy while accurately reproducing the particle distribution. For instance, Lapenta and Brackbill [39] introduced an algorithm for dynamic control of the number of particles in particle-in-cell (PIC) plasma simulation to preserve charge assignment for network nodes without requiring prior knowledge of the particle distribution. Teunissen and Ebert [40] used the k-d tree data structure to facilitate merging particles while maintaining distribution integrity. Similarly, Pfeiffer *et al.* [41] proposed statistical methods for particle split and merge in electromagnetic PIC simulations, emphasizing the conservation of mass, momentum, and energy. Martin and Cambier [42] proposed an octree-based binning approach to maintain distribution fidelity during particle merging. While previous research has often attributed weights to simulator particles, this approach may complicate collision calculations, necessitating the incorporation of new techniques to satisfy conservation laws and control the number of particles after collisions [30,43]. In the SPARTA framework, weights are attributed to cells rather than individual simulator particles, enabling the use of original collision schemes in DSMC without compromising conservation laws during collision.

Besides methods involving particle split and merge, alternative approaches involve replacing the original set of particles with a new set of desired particles. Gorji *et al.* [44] proposed an algorithm based on kernel density estimation to control the number of particles, ensuring the generation of independent statistical samples. This approach involves determining the required number of particles for each cell based on its mass content and sampling particles from the estimated distribution, thereby replacing the existing particles. Apart from DSMC and PIC methods, various strategies for controlling the number of particles have been implemented in other particle-based gas flow simulation methods, such as the ellipsoidal statistical Bhatnagar–Gross–Krook model [45]. Similar control mechanisms have also been proposed for simulating solid-gas flows in which, the solid phase is modelled using the DSMC method [46,47].

The focus of the present work lies in this realm of variable weighting applications. The present work proposes modifications to the SPARTA code, enabling the adjustment of the number of particles per cell (PPC) to a desired value. Acknowledging that achieving absolute precision in adjusting the number of PPC is neither necessary nor feasible, the proposed technique significantly diminishes parameter variations in the number of PPC. To evaluate the physical accuracy of the proposed technique, rarefied gas flow in a short microchannel with a finite pressure ratio and hypersonic argon flow over a cylinder were investigated. Initially, to validate the proposed technique and to determine suitable numerical parameters, the conventional direct simulation Monte Carlo (DSMC) method was employed to simulate gas flow in the microchannel. Subsequently, the "fixed PPC" (FPPC) technique was applied to the same problem, and results were compared with those from the conventional simulation method. Furthermore, two additional case studies were conducted to assess the technique's efficiency in scenarios with notable variations in PPC. The first case study focused on rarefied gas flow in a parallel micromixer, and the second case study examined supersonic rarefied gas flow over a cylinder, evaluating the performance of the proposed technique against the conventional DSMC simulation method.

## 2. The Conventional DSMC Solver

The DSMC simulations were performed using the SPARTA solver [5,37]. SPARTA has been developed and released by Sandia National Laboratories as an open-source software under the GPL (general public licenses) terms. SPARTA can perform both two-dimensional and three-dimensional

DSMC simulations on hierarchical Cartesian grids overlaying the solution domain. The solid objects can be introduced through triangulated surfaces (in the case of three-dimensional simulations) or line segments (in the case of two-dimensional simulations). Cut and split grid cells enable the handling of irregular shapes. Using domain decomposition based on either number of cells, number of particles, or CPU time, SPARTA can run in parallel mode through message-passing techniques [37]. Several modifications have been made by the authors in the source code including a new capability that allows for simulations with a fixed number of particles in all computational cells. This capability, which is the subject of the current study, will be further introduced in section 3.

### 3. Control of the Number of Simulator Particles in Computational Cells

In the conventional implementation of the DSMC method, a constant value is assigned to $F_{num}$ (equivalent number of molecules) across all computational cells, resulting in the average number of particles per cell (PPC) being directly proportional to the product of number density ($n$) and the cell volume ($V_{cell}$). In the current implementation for simulating axisymmetric gas flows in SPARTA software, an additional weight parameter ($w$) has been introduced to regulate the distribution of simulator particles along the radial direction. Hence, each simulator particle represents $w \times F_{num}$ real gas molecules. In the axisymmetric model, this weight parameter is chosen proportionally to the radial position of the cell, *i.e.* the distance from the cell centre to the axis. Alternatively, in SPARTA, the weight parameter can also be chosen in proportion to the cell volume.

The present work focuses on adjusting the number of particles for all computational cells to a target value ($PPC_t$). Unlike many previous methods that assign weight to individual particles, the weight factors in the present work are assigned to cells, ensuring that all particles in a cell share the same weight factor. This approach resolves challenges in satisfying conservation laws and managing the number of particles during collisions. Moreover, unlike the methodology used in prior studies, which involves the real-time determination of particle weights during simulation, this study employs pre-defined weight values that remain constant throughout the simulation. This approach effectively reduces the computational overhead associated with particle count control procedures. To fulfil this, a weight parameter is proposed to be proportional to the number of real gas molecules. To achieve this, an initial estimate of the number of real gas molecules in each cell is required. This estimation is obtained by conducting an initial simulation with a weight parameter of $w = 1$ and a small number of particles (*i.e.* relatively large $F_{num}$). Despite potential inaccuracies in the results due to the limited number of particles, the obtained outcomes serve to estimate the number of real molecules ($N$) or the number of particles in each cell (PPC) in the weightless state. Assuming that the number of simulator particles in the computational cell "*i*" in the initial simulation was calculated as $PPC_i$, the value of the weight parameter ($w_i$) for that cell in the main simulation (with the desired number of particles, $PPC_t$) is determined as follows:

$$w_i = \frac{PPC_i}{PPC_t}. \tag{1}$$

It is noteworthy that the adoption of this technique eliminates the need to modify the parameter $F_{num}$. While the initial simulation needed to achieve the desired goal (*i.e.* adjusting the number of simulator particles to the desired value) may be perceived as incurring an additional simulation time, it serves a dual purpose. This initial simulation can also establish a suitable initial condition for subsequent main executions, mimicking the final steady-state macroscopic condition and thereby reducing the time needed to reach the actual steady-state conditions in the main executions. The technique employed for this purpose is particle scaling, wherein scaling is implemented to set the number of simulator particles to the desired value from the outset of the main execution. Without this technique, progressing from the initial particle count ($PPC_i$) to the desired value ($PPC_t$) can lead to additional simulation time. The scaling parameter, integrated into the SPARTA code, is initially computed for each cell and is defined as follows:

$$s_i = \frac{1}{w_i} = \frac{PPC_t}{PPC_i}. \tag{2}$$

Following this technique, each particle in cell "$i$" should be replaced with $s_i$ particles, all sharing the same characteristics (*i.e.* type, position, velocity vector, and internal energy). Typically, $s_i$ is a decimal number. If the value of $s_i$ is greater than one ($s_i > 1$), its integer part is considered the scale. Subsequently, a random number in the range [0,1] is sampled from a uniform distribution for each initial particle in the cell. If the random number falls in the range of zero to the fractional part of $s_i$, one unit is added to the scale. In cases where $s_i$ is less than one ($s_i < 1$), a random number is sampled for each particle as described earlier. If the selected random number falls in the range of zero to $s_i$, the corresponding particle is retained; otherwise, it is removed. Figure 1 provides an overview of the algorithm employed in the current study for regulating the number of simulator particles in computational cells.

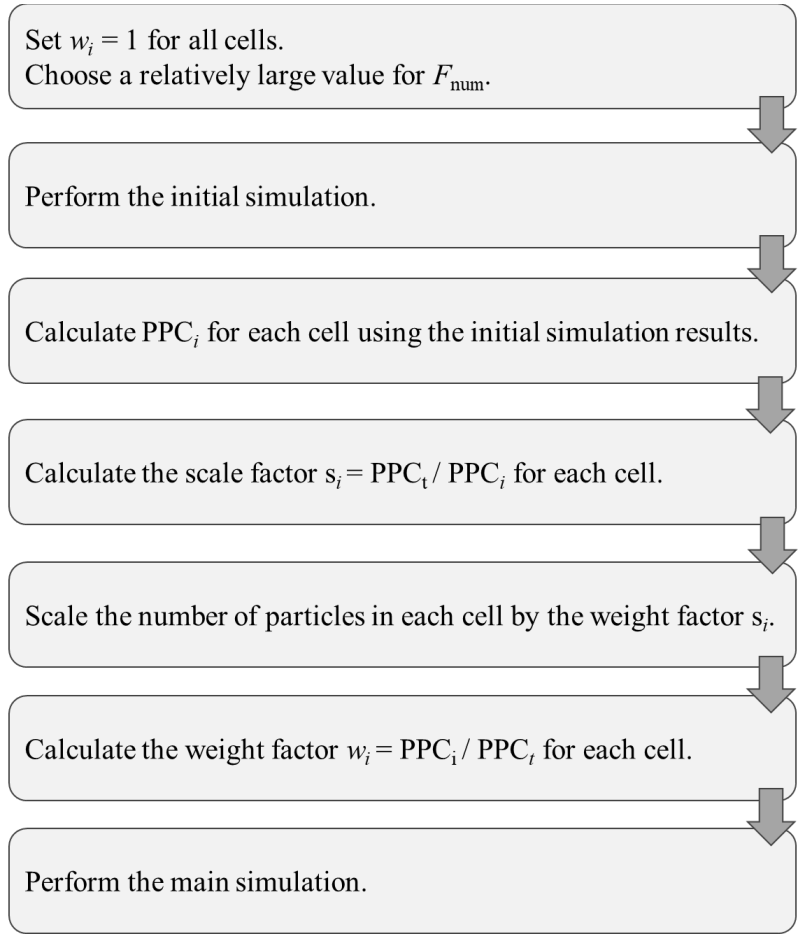

Figure 1. Overview of the algorithm implemented in the current study for regulating the number of simulator particles in computational cells.

## 4. Results and discussion

### 4.1. Validation of the solution method

The accuracy of the FPPC technique in simulations of rarefied gas flow was evaluated by investigating two test cases: rarefied nitrogen flow in a short microchannel with a finite pressure ratio, and hypersonic argon flow over a cylinder.

The simulation of nitrogen gas flow in a microchannel, characterized by a length ($L$) of 2 μm and a height ($H$) of 0.4 μm, was performed using both the conventional and enhanced DSMC solvers. This investigation aimed to assess the accuracy and reliability of the SPARTA solver and the proposed modifications to the SPARTA code. The outlet boundary condition was established at atmospheric pressure, and the inlet-to-outlet pressure ratio ($P_{in} / P_e$) was chosen to be 2.5. The microchannel walls were modelled as fully diffuse with a constant temperature of 300 K. The gas temperature at the inlet was set to 300 K. The Knudsen numbers, determined based on the microchannel height ($H$), at the inlet and outlet correspond to 0.055 and 0.123, respectively.

Considering the symmetry of the flow field along the microchannel centreline, only the upper half of the channel was considered in the model. Both simulations were executed on a uniform $200 \times 60$ Cartesian grid. In the conventional implementation of the solver on a uniform grid, a reduction in the number of particles in each computational cell was observed along the microchannel, with approximately 20 particles per cell at the outlet. Conversely, in the enhanced implementation, this constant particle count was enforced across all cells throughout the entire computational domain.

Previous investigations conducted by Roohi *et al.* [48] and White *et al.* [19] have examined this problem using a conventional implementation of the DSMC method. These studies have presented the dimensionless pressure variation (the ratio of local pressure to the outlet pressure, $P / P_e$) and pressure non-linearity (the relative deviation of pressure variation from a linear distribution, $(P - P_{lin}) / P_{in}$) along the microchannel centreline. Figure 2 shows a comparison between the results obtained from the SPARTA solver and those reported by Roohi *et al.* [48] and White *et al.* [19]. The SPARTA results are obtained using both the conventional implementation and the enhanced implementation (employing the "fixed PPC" (FPPC) technique). Notably, the outcomes of the two implementations align closely, suggesting that the modifications made to the DSMC algorithm to ensure a fixed number of PPC do not yield significant differences in the numerical predictions. Thus, the enhanced implementation can be confidently employed for studying vacuum gas dynamics.

Additionally, the comparison of SPARTA results with those of prior researchers underscores the satisfactory accuracy of SPARTA in both the original and modified modes.

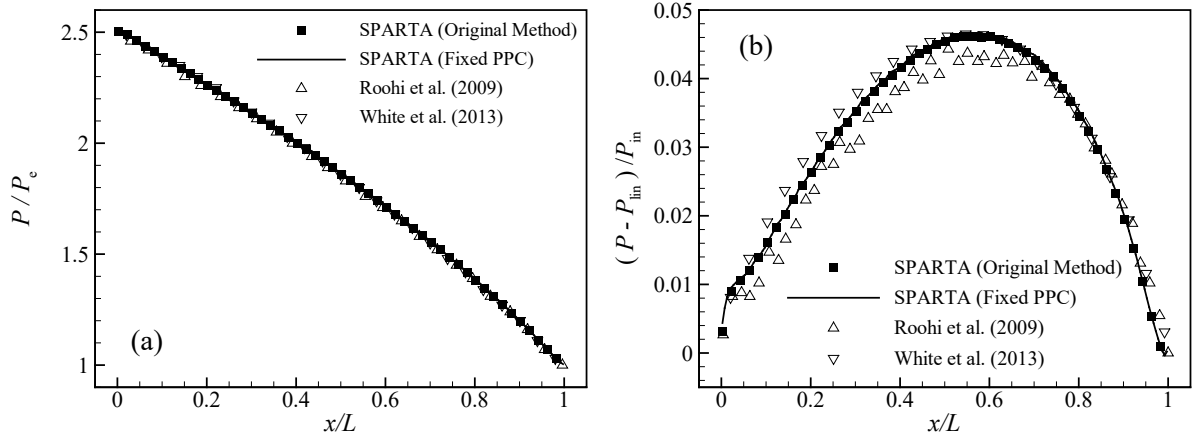

Figure 2. Comparing the outcomes of the SPARTA solver with those presented in references [19] and [48] in the simulation of microchannel flow involves examining: (a) dimensionless pressure variation, and (b) relative non-linearity of pressure variation along the centreline of the microchannel.

A comparison was made between the predictions obtained from DSMC simulations using SPARTA and the analytical solution for gas flow in a microchannel, as shown in Figure 3 and Figure 4. The analytical solution, based on the assumption of isothermal flow and a second-order slip boundary condition, was taken from reference [7]. The non-dimensional pressure distribution along the microchannel, $\hat{P}$, defined as the ratio of the local pressure to the outlet pressure, was computed from the following nonlinear equation:

$$1 - \hat{P}^2 + 12\frac{2-\sigma}{\sigma}\text{Kn}_e(1-\hat{P}) + 12\frac{2-\sigma}{\sigma}\text{Kn}_e^2 \ln(\hat{P}) + B(x-L) = 0, \qquad (3)$$

where $\sigma$ is the wall accommodation coefficient, set to 1.0 in the present study, $L$ is the channel length, and $\text{Kn}_e$ is the Knudsen number at the outlet. The parameter $B$ is computed as follows:

$$B = \frac{\left(1 - P_r^2 + 12\frac{2-\sigma}{\sigma}\text{Kn}_e(1-P_r) + 12\frac{2-\sigma}{\sigma}\text{Kn}_e^2 \ln(P_r)\right)}{L}. \qquad (4)$$

Here, $P_r$ is the inlet-to-outlet pressure ratio. The velocity profile can be calculated as follows:

$$u(y) = \frac{H^2}{2\mu}\frac{dP}{dx}\left[\frac{y^2}{H^2} - \frac{y}{H} + \frac{2-\sigma}{\sigma}(Kn^2 - Kn)\right], \tag{5}$$

where $\mu$ is the dynamic viscosity of the gas, and $H$ is the channel height. Comparisons of pressure variation along the microchannel's centreline and the velocity profile across the mid-length section ($x/L = 0.5$) of the microchannel are presented in Figure 3 and Figure 4, respectively. The results demonstrate a reasonable agreement between the DSMC simulation outcomes and the analytical solutions for both the pressure distribution along the centreline and the velocity profiles across the mid-length section of the microchannel.

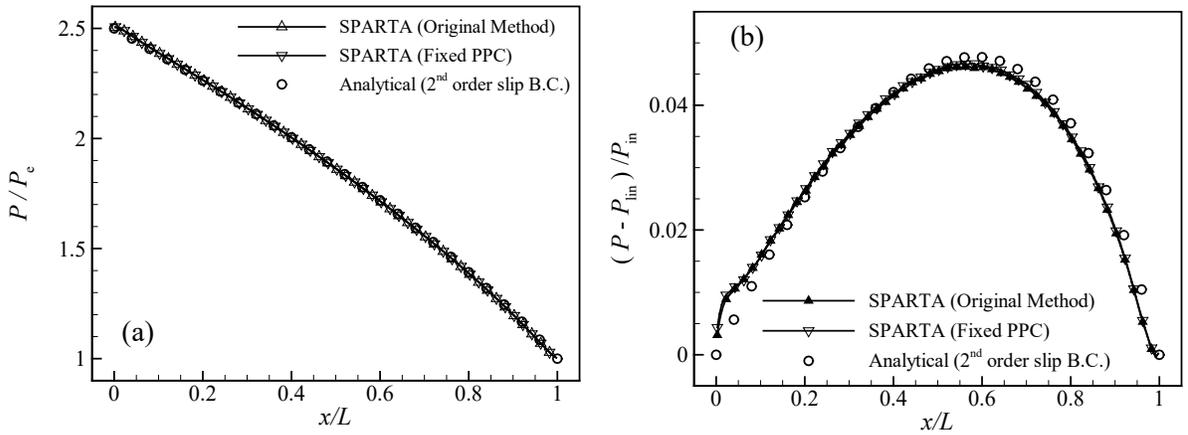

Figure 3. Comparing the outcomes of the SPARTA solver with the analytical solution using the second-order slip boundary condition, examining: (a) dimensionless pressure variation, and (b) relative non-linearity of pressure variation along the centreline of the microchannel.

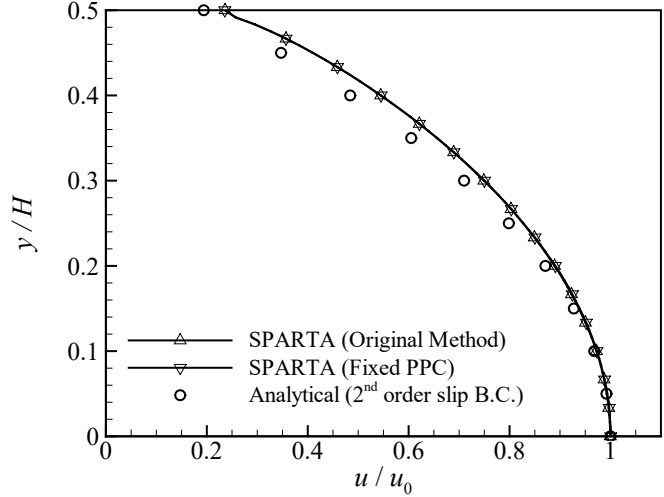

Figure 4. Comparing the outcomes of the SPARTA solver with the analytical solution using the second-order slip boundary condition in predicting the velocity profile across the mid-length section ($x/L = 0.5$) of the microchannel.

As the second validation test case, external hypersonic argon flow with a Mach number of 10 over a 12-inch (~30.48 cm) cylinder was considered. This problem has been studied by several researchers (see for instance, [49-51]). The flow was simulated using the SPARTA solver, and both the conventional DSMC implementation and the enhanced implementation with the FPPC technique were employed. In this problem, the free stream was characterised by a temperature of 200 K, a number density of $4.247 \times 10^{20}$ m$^{-3}$, and a velocity of 2634.1 m s$^{-1}$. The cylinder surface was modelled as a fully diffuse surface with a temperature of 500 K. The parameters of the VHS collision model for argon gas were taken from [49]. Due to the symmetric configuration of the fluid flow, only the upper half of the physical domain was used as the computational domain. A uniform grid with 1500×900 cells was employed, resulting in a cell size smaller than 1/4 of the molecular mean free path ($\lambda$) in the free stream region. The time step size was set to $1 \times 10^{-7}$ s, which is approximately 85 times smaller than the mean collision time and guarantees sufficient time (on average two time-steps) for particles to cross the width of a computational cell. When the conventional DSMC implementation was used, the value of $F_{num}$ was chosen such that each computational cell in the free stream region contained about 10 simulator particles. When the enhanced implementation with the FPPC technique was used, the value of PPC was set to 10.

Table 1 presents a comparison of the predictions obtained from DSMC simulations using both the conventional DSMC implementation and the enhanced implementation with the FPPC technique. The parameters examined are the drag force exerted on a full cylinder and the peak heat flux. For validation, these results are compared with those reported in previous studies [49-51]. The results indicate that the predicted values for drag force are consistent across different methodologies. Both the conventional DSMC implementation and the FPPC-enhanced implementation report a drag force of 39.8 N. In comparison, Lofthouse [49] reports a slightly higher drag force of 40.02 N, while Lo et al. [50] report 39.89 N. The peak heat flux values show similar trends; the peak value of heat flux obtained using the conventional DSMC implementation is 38.60 kW m$^{-2}$, while the FPPC-enhanced implementation predicts a slightly higher value of 39.13 kW m$^{-2}$. This peak heat flux value is consistent with the findings of Lofthouse [49], and is in close agreement with the results of Goshayeshi et al. [51] and Lo et al. [50].

Table 1. Comparison of drag force and peak heat flux magnitudes from DSMC simulations, using both the conventional implementation and the FPPC-enhanced implementation, for hypersonic argon flow over a cylinder (Ma = 10), with data reported in the literature [49-51].

|  | Magnitude of drag force [N] | Magnitude of peak heat flux ($\theta = 180°$) [kW m$^{-2}$] |
| --- | --- | --- |
| SPARTA (Conventional implementation) | 39.8 | 38.60 |
| SPARTA (FPPC) | 39.8 | 39.13 |
| Lofthouse [49] | 40.02 | 39.13 |
| Lo et al. [50] (VTS+TAS) | 39.89 | 38.66 |
| Goshayeshi et al. [51] | N.A. | ~ 38 |

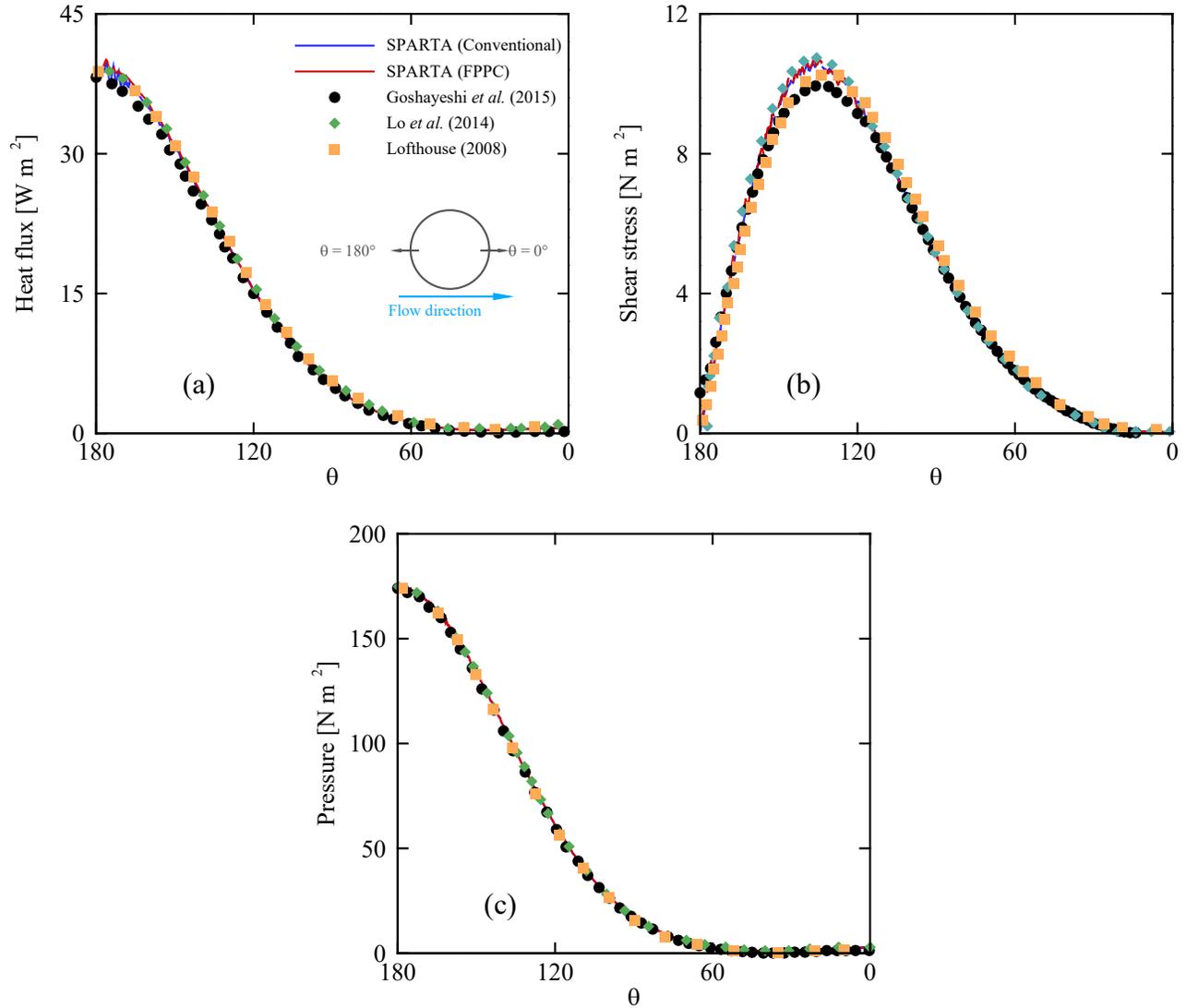

Figure 5. Comparison of the distributions of (a) heat flux, (b) shear stress, and (c) pressure obtained from DSMC simulations using both conventional and FPPC-enhanced implementations for hypersonic argon flow over a cylinder (Ma = 10). The simulation results are compared with data reported by Lofthouse [49], Lo *et al*. [50], and Goshayeshi *et al*. [51].

Figure 5 shows the distributions of heat flux, shear stress and pressure obtained from DSMC simulations using both the conventional DSMC implementation and the FPPC-enhanced implementation for hypersonic argon flow over a cylinder. These simulation results are compared with the data reported in previous studies [49-51]. The results demonstrate good agreement between the conventional and FPPC-enhanced DSMC implementations. The DSMC simulations, both conventional and FPPC-enhanced, agree closely with the results reported in previous studies, indicating the reliability and accuracy of the present DSMC simulations in predicting hypersonic rarefied gas flows over bluff bodies. This validation also confirms that the enhancements introduced by the FPPC technique do not adversely affect the accuracy of the simulations, while potentially offering computational advantages.

To assess the efficacy of the enhanced implementation (*i.e.* the FPPC implementation), two rarefied gas flow problems were studied. Subsequent sections present a comparative analysis between the outcomes obtained using the enhanced implementation and those generated by the conventional implementation. Additionally, an evaluation of the computational efficiency improvement is conducted through a comparison of the CPU times associated with the simulations. Each simulation was executed on a desktop computer with 12 cores (Intel Xeon E5530 @ 2.40 GHz) and 11 GB of physical memory (RAM).

### 4.2. Micromixer Flow

In numerical investigations of fluid flow, refining the computational grid size in regions with strong local gradients, such as mixing layers, near-wall regions, and shock waves is often critical for accurate prediction of variable changes. In DSMC simulations, the value of $F_{num}$ is generally a constant. Hence, employing a fine grid size leads to fewer particles in computational cells, which may not be favourable due to the increase in statistical errors. If a minimum particle criterion is imposed in the conventional implementations of the DSMC method, the value of $F_{num}$ needs to be reduced to meet this criterion, resulting in an increase in the number of particles in all cells (*i.e.* the total number of simulator particles) and computational costs. However, implementing the proposed FPPC technique can mitigate this increase in the computational costs.

An example illustrating the impact of strong local gradients on rarefied conditions and molecular diffusion is evident in the mixing of two gas species in a parallel micromixer, as depicted in Figure 6. In vicinity of the splitter's end, there is an obvious increase in the gradients of mixture constituent mole fractions, leading to augmented diffusive mass fluxes. This intense local gradient induces significant rarefied conditions, consequently diminishing the effectiveness of the molecular diffusion mechanism. This reduction of molecular diffusion is reflected in a reduction of the equivalent diffusion coefficient compared to its value under quasi-equilibrium conditions, without extreme gradients [52]. A thorough investigation into the reduction of molecular diffusion near the splitter's end, characterized by a severe mole fraction gradient, requires the utilization of relatively fine grid sizes, a substantial number of molecules, which can consequently result in an augmented computational time.

Darbandi and Sabouri [52] employed the DSMC method to investigate the mixing of two gas species in a parallel micromixer with specular walls. Their focus was on exploring the influence of concentration gradients on molecular diffusion. The outcomes reported by Darbandi and Sabouri [52] indicate a significant increase in the gradients of mole fractions and, consequently, diffusive mass fluxes near the splitter's end. The present study demonstrates that local refinement of the computational grid near the splitter's end, coupled with the proposed FPPC technique, enables accurate prediction of variations in high-gradient field quantities at a reduced computational cost compared to conventional implementations of the DSMC method.

The investigated problem involves the mixing of gases A and B in a parallel mixer with characteristics analogous to the variable soft sphere (VSS) model for nitrogen gas from reference [1]. The mixer's geometry is shown in Figure 6, where gases A and B are introduced through lower and upper inlet channels, respectively. Prior to the mixing channel, a splitter plate maintains the separation of the two streams. The walls of both the mixer and splitter are treated as fully specular. Uniform flow conditions at the entrance, characterized by uniform velocity, temperature, and density distributions, are assumed. At the outlet, a constant pressure condition matching the inlet pressure is set, establishing horizontal uniform flow throughout the mixer. The Knudsen number, calculated using the height of each inlet channel as the characteristic length, is 0.1. Inlet flow conditions include a temperature of 300 K and a velocity of 150 m s$^{-1}$.

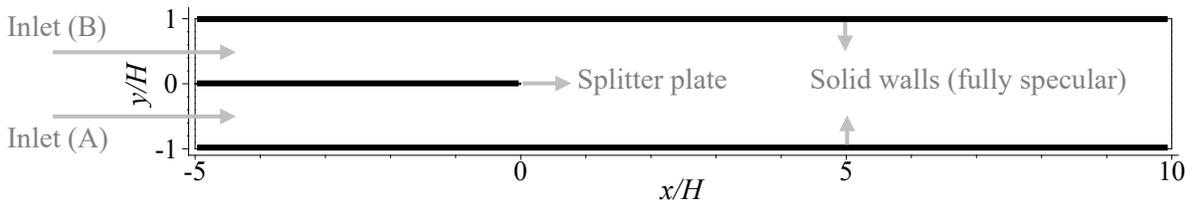

Figure 6. Schematic of the parallel mixer studied in the present work.

Simulations were conducted on three Cartesian grids with dimensions of 75×20, 150×40, and 300×80 uniformly spaced computational cells. In each case, the parameter $F_{num}$ was adjusted to maintain an average of around twenty particles per cell. Figure 7 shows the variations in the mole fraction of gas species A across different sections of the mixer predicted using different grid sizes. The graph corresponding to the end of the splitter's cross-section ($x / H = 0$) indicates a significant mole fraction gradient in the middle of this section, particularly near the end of the splitter. Notably, differences in results become apparent in this region, emphasizing the need for a fine grid to accurately calculate the local gradient of flow variables. Moving away from the splitter's end, both the mole fraction gradient and the sensitivity to grid size decrease.

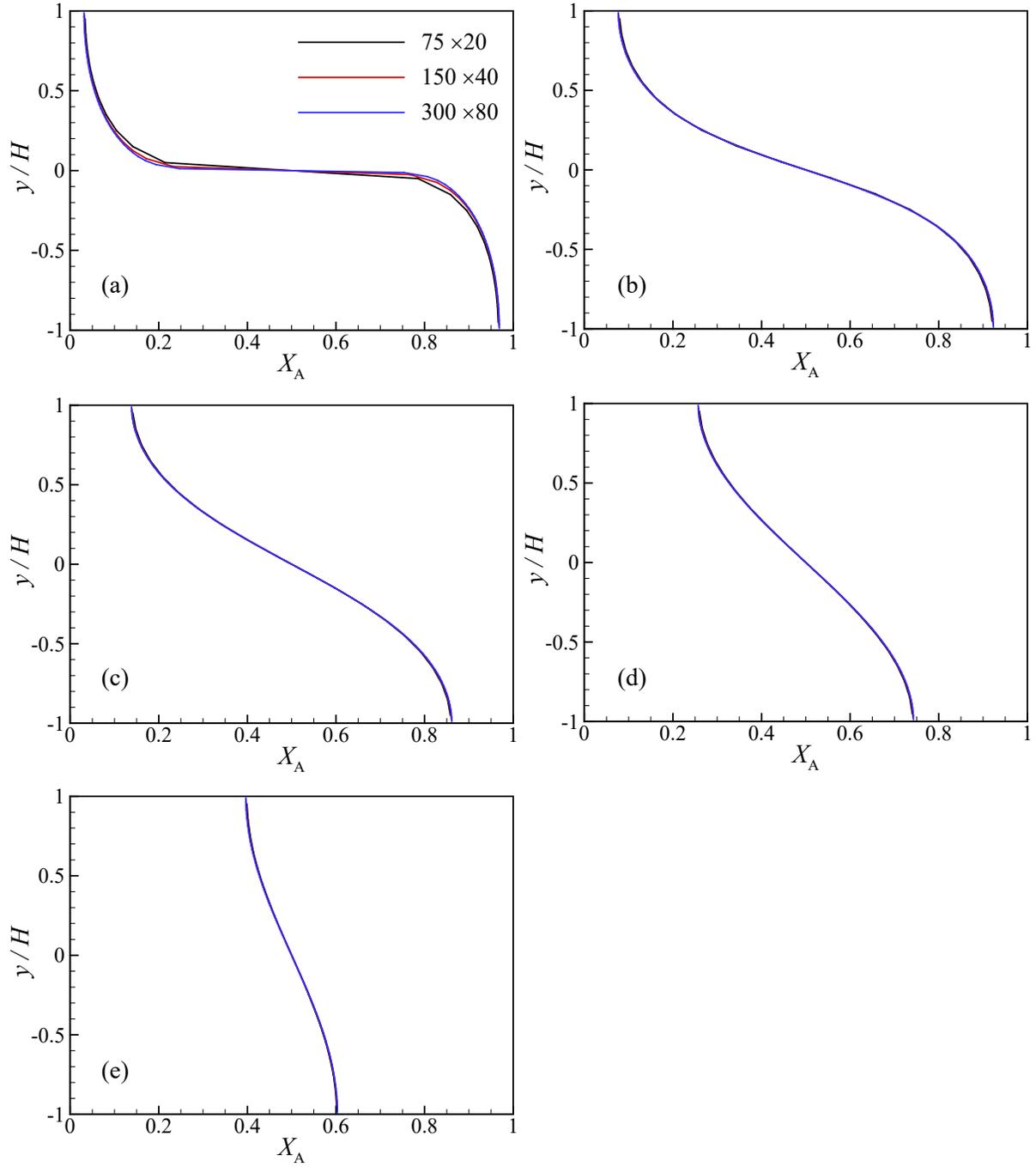

Figure 7. Effect of computational grid size on the numerical predictions of the mole fraction of gas species A across different sections of the mixer: (a) $x/H = 0$, (b) $x/H = 0.4$, (c) $x/H = 0.8$, (d) $x/H = 1.6$, and (e) $x/H = 3.2$.

Figure 8 and Figure 9 show the variations in the *x* and *y* components of the dimensionless diffusive mass flux of gas species A across different sections of the mixer, respectively. The dimensionless diffusive mass flux, $J_A^*$, is defined as follows:

$$J_A^* = \frac{H J_A}{\rho D_{AB}}, \qquad (6)$$

where, *H* is the height of each of the inlet channels, $\rho$ is the density of the mixture, and $D_{AB}$ is the binary diffusion coefficient. In regions close to the end of the splitter, a sharp gradient in the mole fraction results in a substantial increase in the diffusive flux. Significant disparities among the results obtained from different grid sizes are notable in the proximity of the splitter's end, with these differences decreasing as moving away from this point.

Based on the findings presented in Figure 8 and Figure 9, implementing local grid refinement in regions with significant gradients, particularly close to the end of the splitter, and applying the FPPC technique, appears feasible to achieve a similar (or better) accuracy while reducing the computational costs. The grid with 75×20 cells was selected, and grid refinements based on the octree refinement approach were executed in two different regions. In the first step, cells located in the region $-1 \leq x/H \leq 2$ were subdivided into four cells. Subsequently, a second refinement was applied to cells located in the region $-0.5 \leq x/H \leq 0.5$. Two simulations were conducted on this grid. In the first simulation without utilising the FPPC technique, the average particle count per cell was proportionate to the cell volume due to uniform density distribution. By adjusting the value of $F_{num}$, the average number of particles in the smallest cells was set to twenty. In this case, larger cells contained more particles. Despite using this non-uniform grid (with local grid refinement), the total number of particles closely approximated that of the simulation using the 300×80 uniform grid. In the second simulation, employing the FPPC technique confirmed that the number of particles in all cells of the non-uniform grid remained approximately twenty.

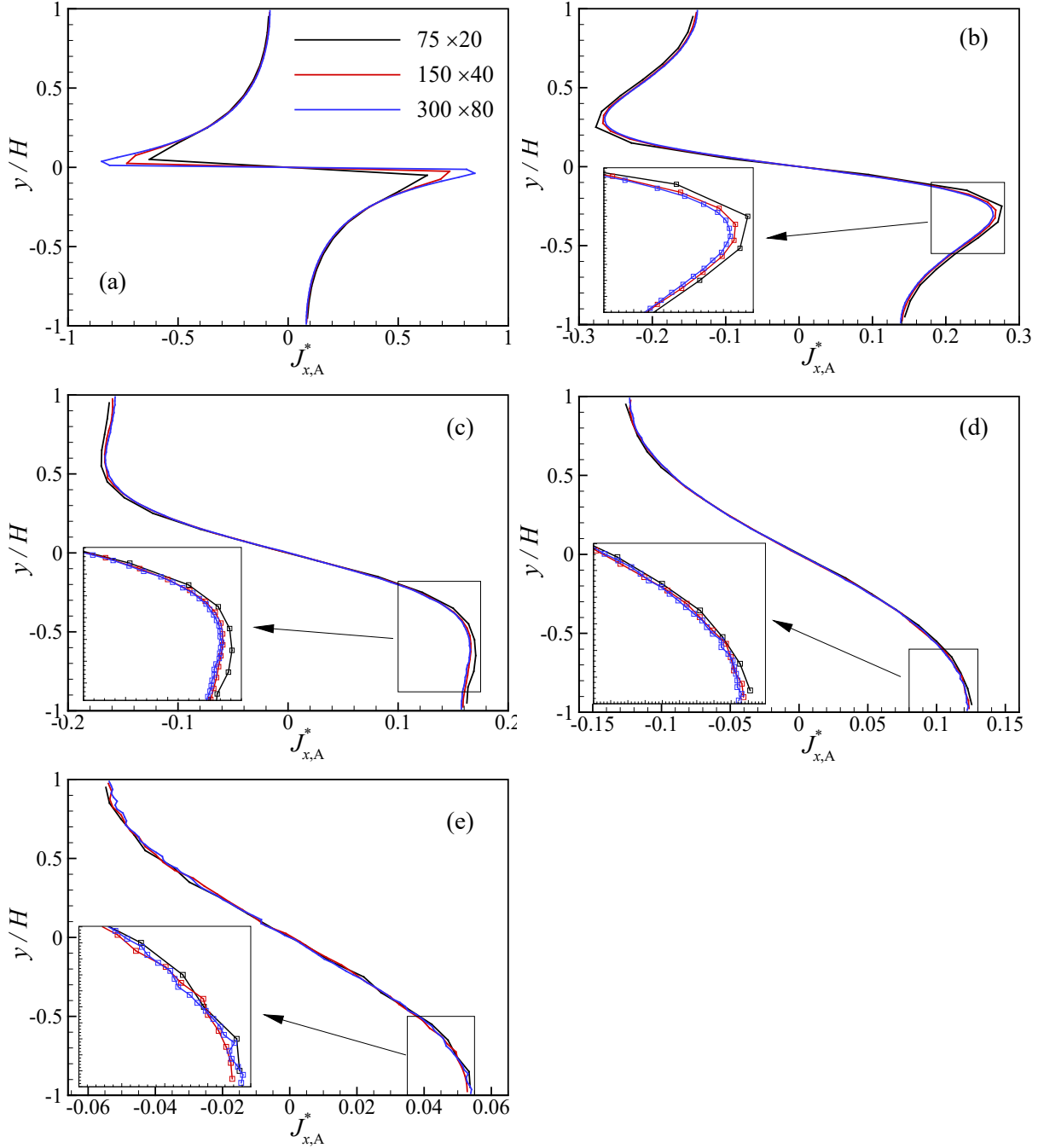

Figure 8. Effect of computational grid size on the variations of the dimensionless diffusive mass flux of gas species A in the +x direction across different sections of the mixer: (a) $x/H = 0$, (b) $x/H = 0.4$, (c) $x/H = 0.8$, (d) $x/H = 1.6$, and (e) $x/H = 3.2$.

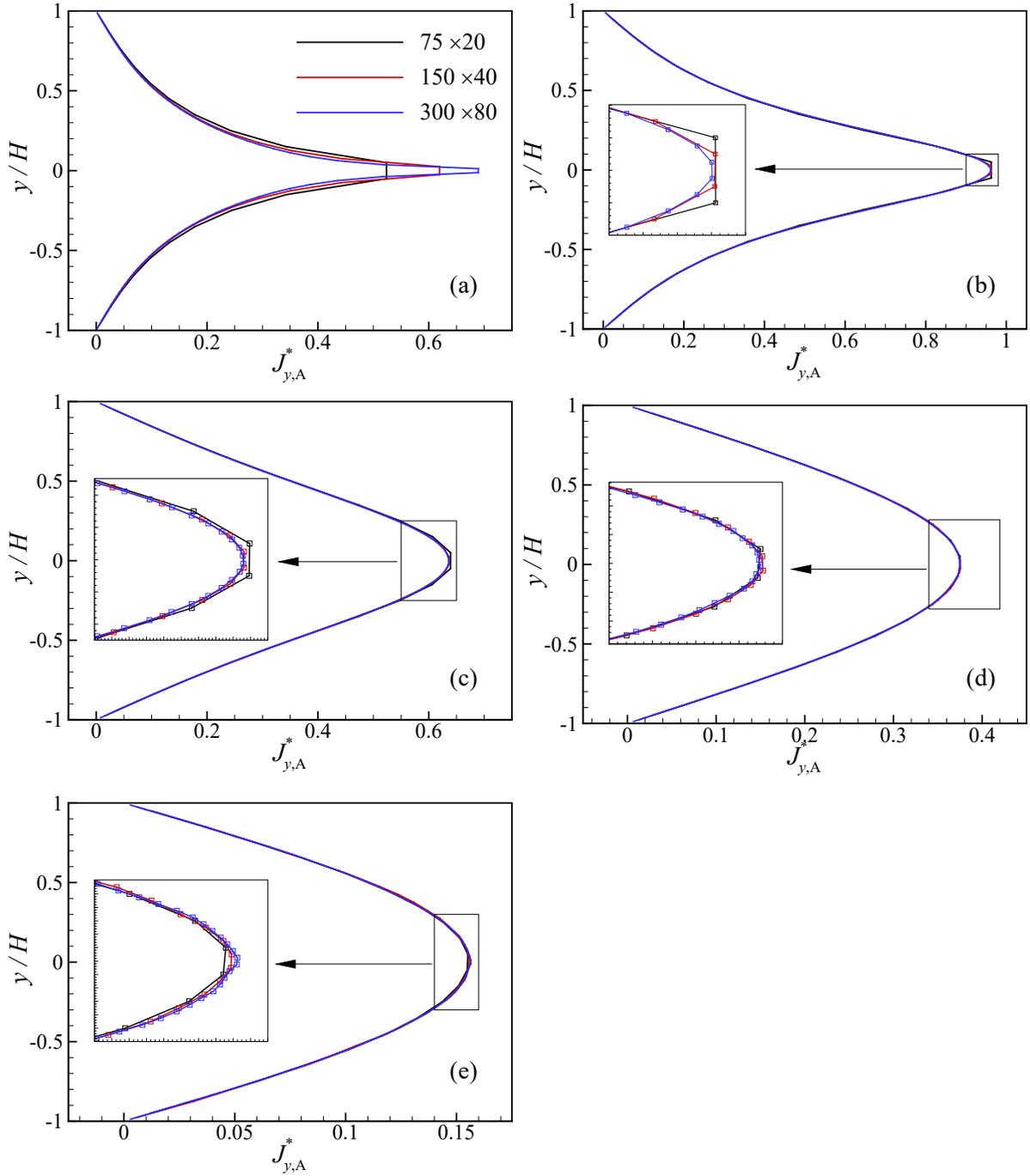

Figure 9. Effect of computational grid size on the variations of the dimensionless diffusive mass flux of gas species A in the +$y$ direction across different sections of the mixer: (a) $x/H = 0$, (b) $x/H = 0.4$, (c) $x/H = 0.8$, (d) $x/H = 1.6$, and (e) $x/H = 3.2$.

To assess the accuracy and computational efficiency of the FPPC technique, comparisons were made among the results obtained from simulations conducted using the uniform 300×80 grid, the locally refined grid without implementing the FPPC technique, and the locally refined grid utilizing the FPPC technique. Figure 10 to Figure 12 depict the distributions of mole fraction, as well as the $x$ and $y$ components of the dimensionless diffusive mass flux of gas species A for various sections of the mixer. A high level of agreement is observed across the results of the three simulations. The agreement between the result of the uniform fine grid and the locally refined grid without using FPPC technique indicates the capability of a non-uniform grid with local refinement in gradient-rich regions to return comparable results to a finely uniform grid. Moreover, the agreement of the results obtained from simulations with the FPPC technique with those obtained without this technique demonstrates the accuracy of the FPPC technique.

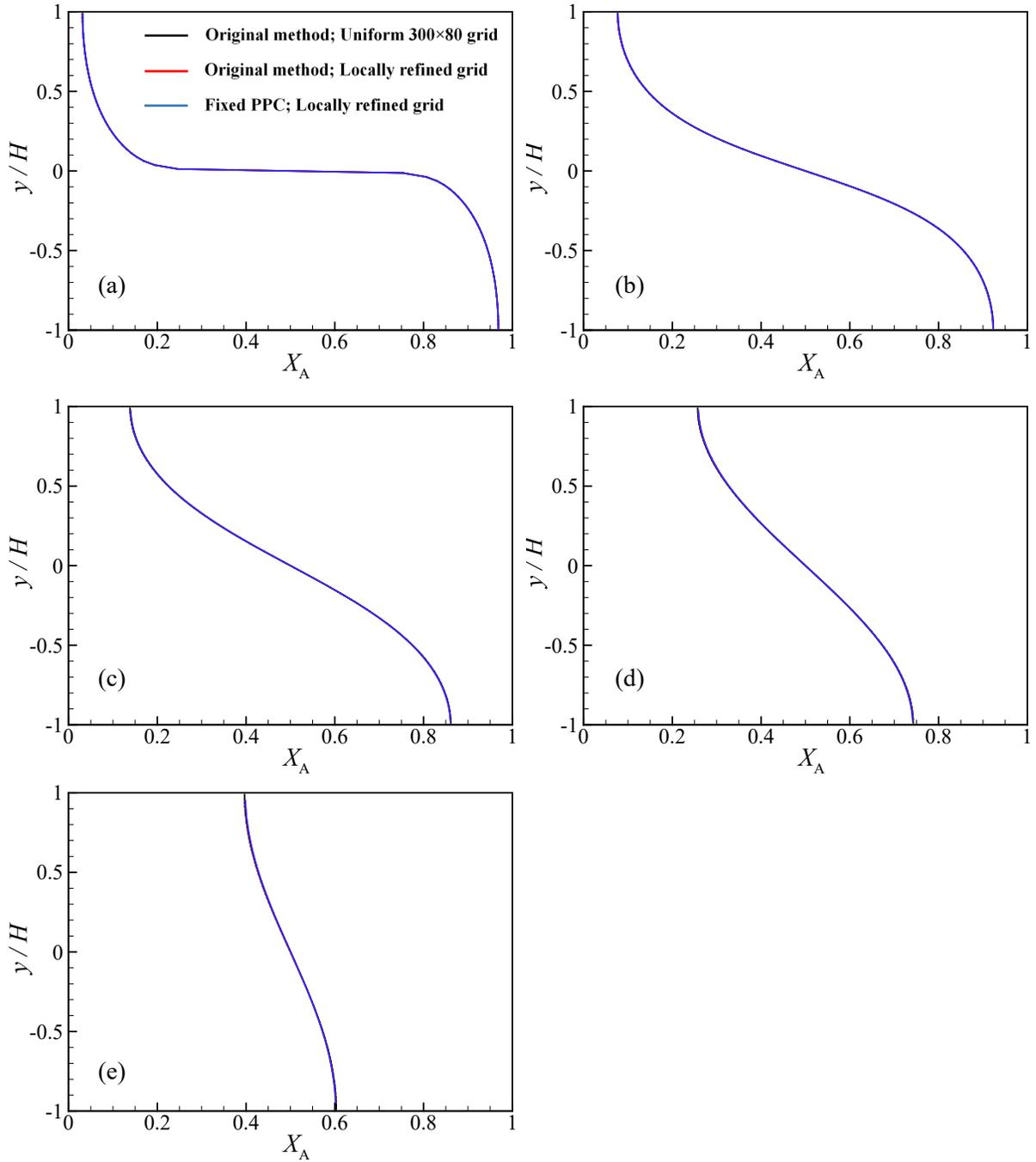

Figure 10. Effect of different simulation approaches, including the conventional method on a uniform 300×80 grid, the conventional method on a locally refined grid, and the FPCC method on a locally refined grid, on the variations of the mole fraction of species A across different sections of the mixer: (a) $x/H = 0$, (b) $x/H = 0.4$, (c) $x/H = 0.8$, (d) $x/H = 1.6$, and (e) $x/H = 3.2$.

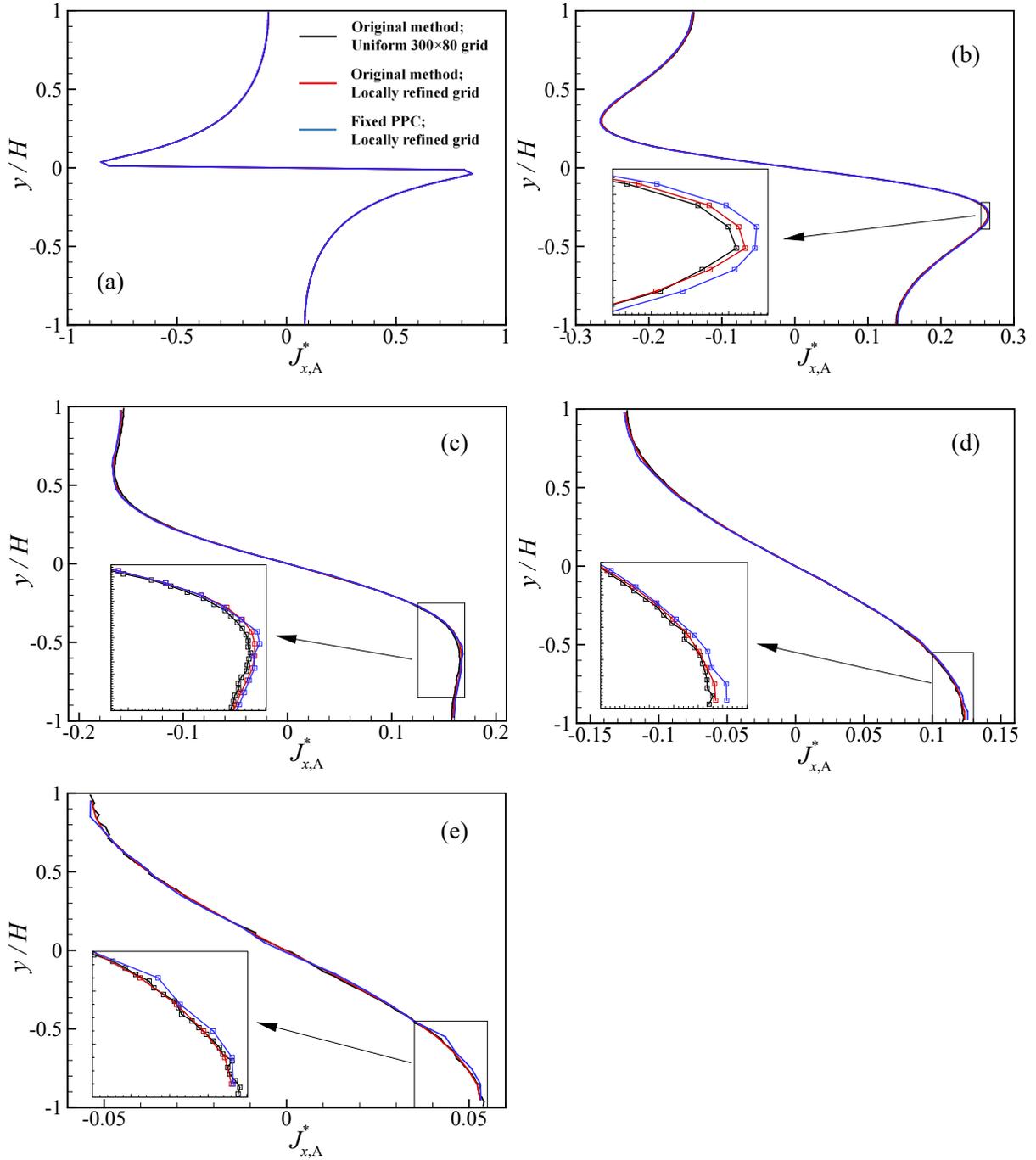

Figure 11. Effect of different simulation approaches, including the conventional method on a uniform 300×80 grid, the conventional method on a locally refined grid, and the FPCC method on a locally refined grid, on the variations of the dimensionless diffusive mass flux of gas species A in the +$x$ direction across different sections of the mixer: (a) $x/H = 0$, (b) $x/H = 0.4$, (c) $x/H = 0.8$, (d) $x/H = 1.6$, and (e) $x/H = 3.2$.

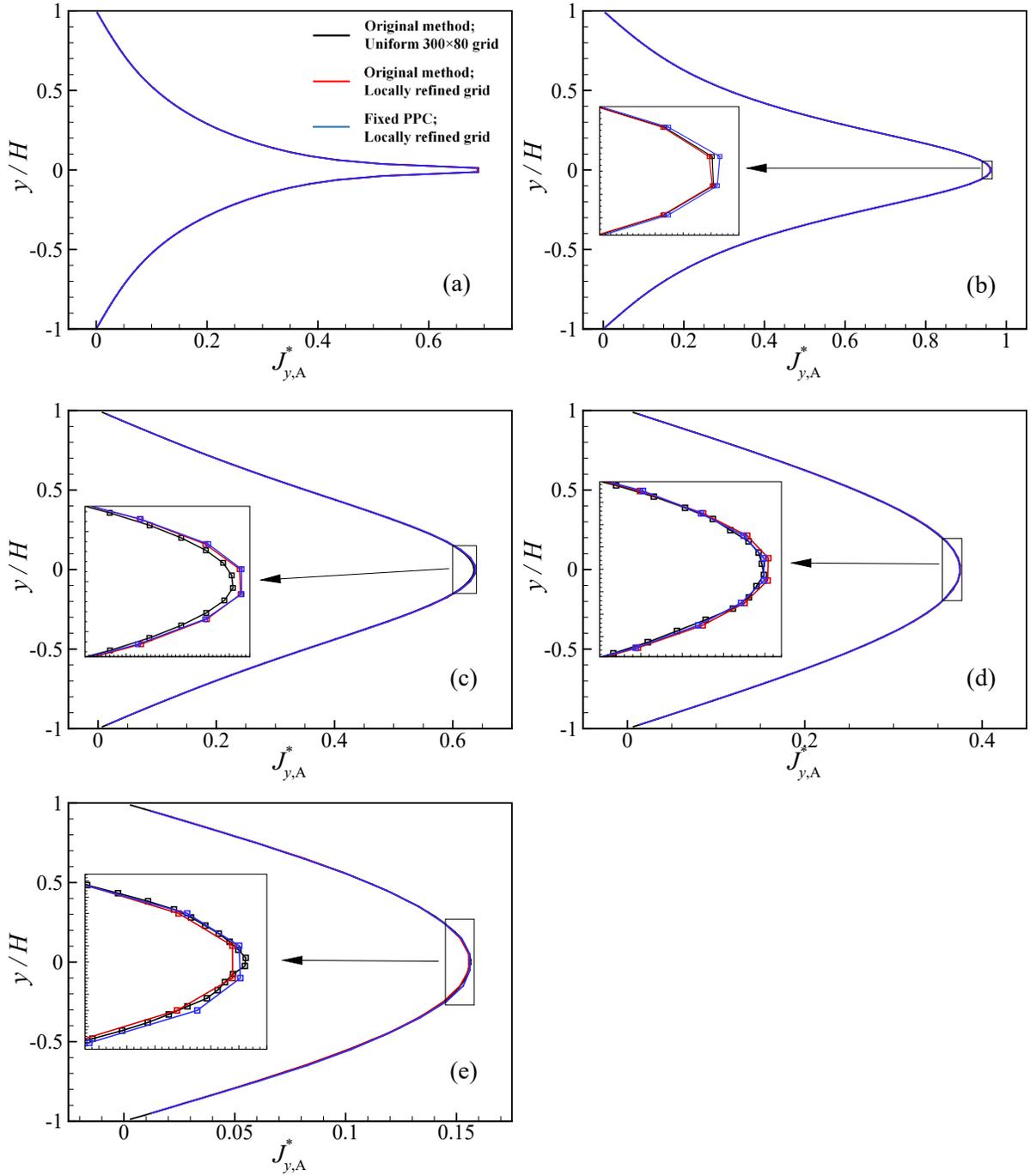

Figure 12. Effect of different simulation approaches, including the conventional method on a uniform 300×80 grid, the conventional method on a locally refined grid, and the FPCC method on a locally refined grid, on the variations of the dimensionless diffusive mass flux of gas species A in the +$y$ direction across different sections of the mixer: (a) $x/H = 0$, (b) $x/H = 0.4$, (c) $x/H = 0.8$, (d) $x/H = 1.6$, and (e) $x/H = 3.2$.
28

To further evaluate the accuracy of the FPPC technique, the relative difference between the results obtained from simulations using the FPPC technique and those using the conventional DSMC method was calculated for the mole fraction distribution ($X_A$) and the components of the dimensionless diffusive mass flux of species A (*i.e.* $J_{A,x}$, and $J_{A,y}$). The relative difference was determined using the following formula:

$$\mathcal{D} = \frac{\xi_{FPPC} - \xi_{Conventional}}{\xi_{Conventional}} \cdot 100, \tag{7}$$

where, $\xi$ represents the quantity of interest (*i.e.* $X_A$, $J_{A,x}$, and $J_{A,y}$). Figure 13 shows that the relative difference between the simulation results using the FPPC technique and the conventional method is generally less than 2% across all sections. The results indicate that the relative difference between model predictions increases as the micromixer outlet is approached. Notably, the relative difference in predicting dimensionless diffusive mass flux along the *x*-direction becomes pronounced due to significantly smaller values along this direction compared to those along the *y*-direction. It is important to note that DSMC simulation results inherently include statistical fluctuations. Consequently, when the value of a quantity approaches zero, these fluctuations can be significant compared to the actual value of the quantity. Thus, the relative error calculated under these conditions is primarily due to statistical error rather than a systematic error in the computational method.

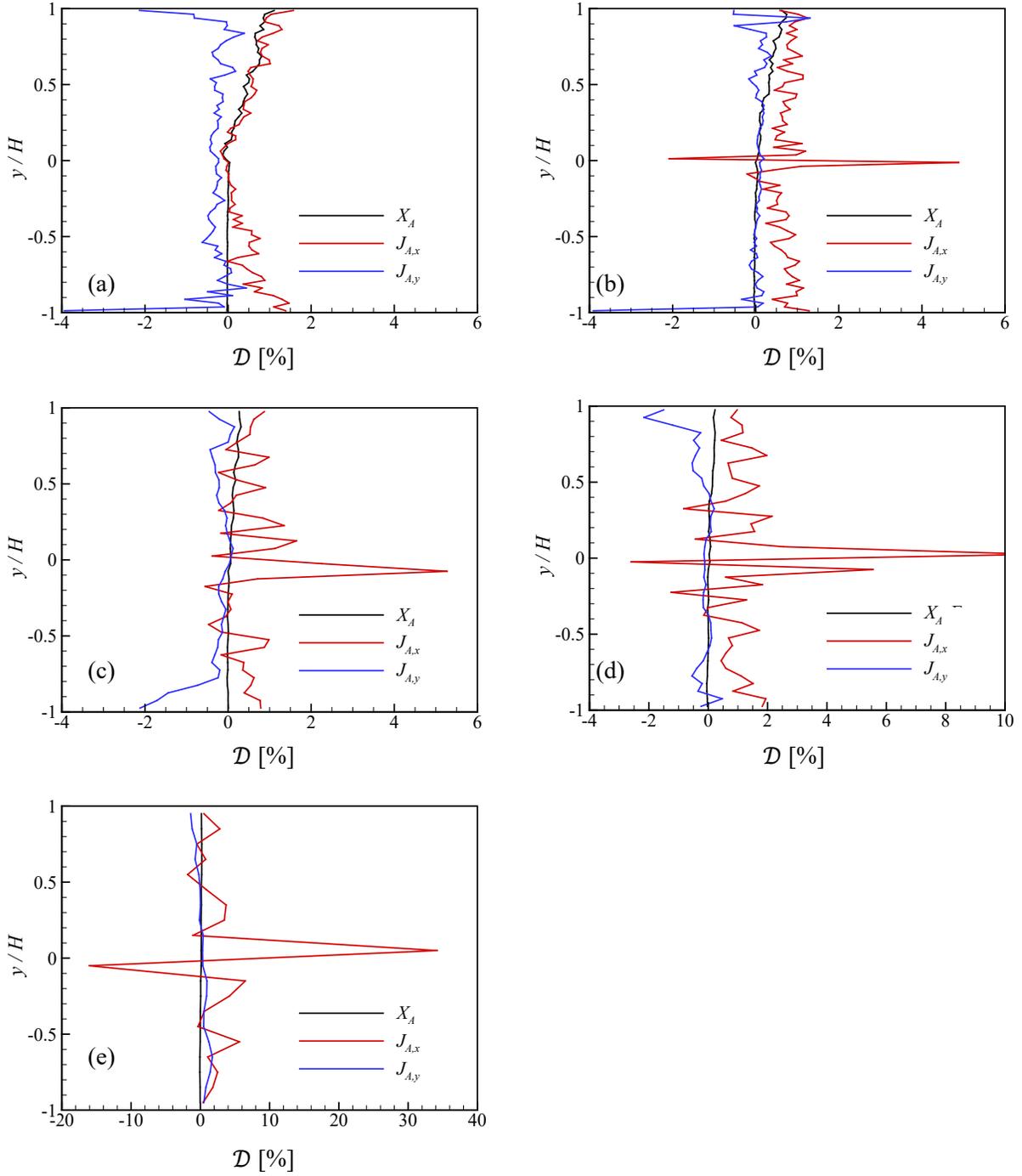

Figure 13. Relative difference between the results obtained from simulations using the FPPC technique and those using the conventional DSMC method for the mole fraction distribution ($X_A$) and the components of the dimensionless diffusive mass flux of species A ($J_{A,x}$, and $J_{A,y}$): (a) $x/H = 0$, (b) $x/H = 0.4$, (c) $x/H = 0.8$, (d) $x/H = 1.6$, and (e) $x/H = 3.2$.

To evaluate the computational efficiency of different approaches, the CPU time required to complete $10^7$ time-steps was measured, and the results are summarized in Table 2. The findings indicate that the simulations were completed in 54.4 hours, 81.8 hours, and 4.7 hours using the fine uniform grid without the FPPC method, the locally refined grid without the FPPC method, and the locally refined grid with the FPPC method, respectively. These simulations included both the initial and final stages of the simulation process. The initial stage includes time steps prior to the initiation of the sampling process. For simulations employing the FPPC method, this stage also involves the preliminary simulation needed to calculate the weight factors. The final stage includes the time steps following the initiation of the sampling process, incorporating the computational costs of molecular dynamics computations, such as motion, sorting, particle tracking, collision, and sampling. For a more reasonable comparison, it is recommended to focus on the computational costs incurred during the final stage.

Considering the proportional relationship between computation time and the number of simulation steps in DSMC simulations, the average CPU time per time step serves as a suitable metric for comparing computational efficiency. Table 2 presents the average CPU times per time step and the corresponding average number of particles for the three test cases. As anticipated, local grid refinement without employing the FPPC technique did not cause a significant change in the number of particles. The slight difference between the number of particles in the cases of the uniform fine grid and the locally refined grid without the FPPC technique is attributed to inherent fluctuations in DSMC simulations. However, the local grid refinement (without implementing the FPPC technique) resulted in approximately a 45% increase in computation time. This increase is likely attributable to the increased computational cost of the particle tracking process. In a uniform Cartesian grid, determining the cell for each particle can be swiftly accomplished with a few straightforward calculations. In contrast, in a non-uniform grid, determining the cell for each particle requires more computations. It is essential to note that the reported increase in computational cost may not be universally applicable to all DSMC solvers, as it could vary depending on the grid structure or algorithms employed by each solver.

Implementing the FPPC technique resulted in a substantial decrease in both the total number of particles and calculation time. The application of this technique led to a 6.6-fold reduction in the number of particles, reducing the calculation time by approximately 16.6 times compared to the scenario without FPPC on a non-uniform grid and around 11.5 times compared to the case with a fine uniform grid. Considering the influence of the number of particles on the computational cost per time step, the CPU time per particle for each time step was computed and presented in Table 2. The application of a non-uniform grid without the FPPC technique increased the CPU time by about 45%. In contrast, employing the FPPC technique reduced the CPU time per particle by about 1.7 times compared to the uniform grid mode and approximately 2.5 times compared to the non-uniform grid mode without the FPPC technique. This notable reduction is likely attributed to the decreased cost of collision calculations.

Table 2. The effect of three different approaches, including the uniform fine grid, the locally refined grid without the FPPC idea, and the locally refined grid using the FPPC idea on the computational cost in the simulation of mixer flow.

| Case | Number of Particles | CPU time for a complete simulation (h) | CPU time per time step (ms) | CPU time per time step per particle (μs) |
|---|---|---|---|---|
| Uniform grid 300×80 without FPPC | 479825 | 54.4 | 19.84 | 41.34 |
| Locally refined grid without FPPC | 479807 | 81.8 | 28.79 | 60.00 |
| Locally refined grid with FPPC | 71975 | 4.7 | 1.73 | 23.98 |

## 4.3. Supersonic airflow over a cylinder

Grid refinement in specific regions of the computational domain may be needed in simulations of complex internal and external fluid flows. SPARTA utilises a Cartesian grid with an octree grid structure for local grid refinement. This grid refinement leads to a reduction in the number of particles in each refined cell. As the second example, a two-dimensional supersonic rarefied gas flow around a cylinder was simulated. The simulation considered airflow at a Mach number of 2 and a Knudsen number of 0.025 with the cylinder's diameter as the characteristic length scale. Figure 14 shows the computational domain, with length scales normalised to the cylinder diameter, $d$. The computational domain was designed to ensure uniform flow conditions at the inlet boundaries and to prevent the shock wave from crossing its boundaries. Considering the problem's symmetry, only the upper half of the flow field was used in the simulations.

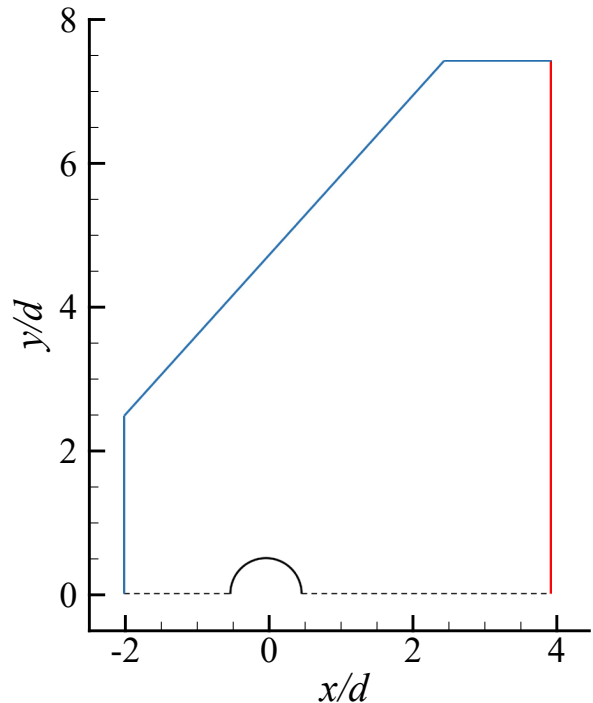

Figure 14. Schematic of the computational domain employed in the present work to study rarefied gas flow around a cylinder; Blue lines indicate the free flow inlet boundary, the red line shows the outlet boundary, the black solid line shows the cylinder solid wall, and the black dashed lines indicates the symmetry boundary.

Four simulations were conducted using uniform Cartesian grids with 120×150, 240×300, 480×600, and 960×1200 computational cells. In each simulation, the value of $F_{num}$ was adjusted to ensure an average of approximately twenty particles per cell in the free flow region. To assess grid quality, the two components of the force acting on the semi-cylinder were computed for different grid resolutions. Notably, the $x$ component of this force represents half of the drag on the full cylinder, while the $y$ component is compensated by the force acting on the lower half of the cylinder, resulting in a net lift force of zero for the complete cylinder. Table 3 presents the dimensionless values of these force components for the four different grids, utilizing the following dimensionless scaling relations:

$$c_{x,\text{HM}} = 4 \frac{F_{x,\text{half model}}}{\rho_\infty u_\infty^2 d}, \tag{8}$$

$$c_{y,\text{HM}} = -2 \frac{F_{y,\text{half model}}}{\rho_\infty u_\infty^2 d}, \tag{9}$$

where, $\rho_\infty$ and $u_\infty$ are the free stream density and velocity, respectively. The dimensionless force in the $x$ direction corresponds to the drag coefficient for the complete cylinder. The results indicate that refining the grid size from 480×600 to 960×1200 does not have a noticeable impact on the predicted force coefficients. Therefore, the 480×600 grid resolution is fine enough to obtain reliable results.

Table 3. The effect of computational grid size on the predicted dimensionless force coefficients.

| Grid | $c_{x,\text{HM}}$ | Relative difference w.r.t the finest grid | $c_{y,\text{HM}}$ | Relative difference w.r.t the finest grid |
|---|---|---|---|---|
| 120×150 | 1.514 | 1.88% | 0.553 | 2.60% |
| 240×300 | 1.492 | 0.40% | 0.542 | 0.56% |
| 480×600 | 1.486 | 0.00% | 0.539 | 0.00% |
| 960×1200 | 1.486 | - | 0.539 | - |

While the 480×600 grid appears suitable for the present problem, computational cell sizes in certain regions in the computational domain may be smaller than needed. To investigate this, the distributions of *x*-Mach number (the ratio of the *x*-component of the velocity vector to the speed of sound), pressure ratio to free flow pressure, and temperature ratio to free flow temperature along the symmetry boundary ($y = 0$) and across the vertical line passing through the centre ($x = 0$) are compared for three distinct grid sizes, and the results are shown in Figure 15 and Figure 16.

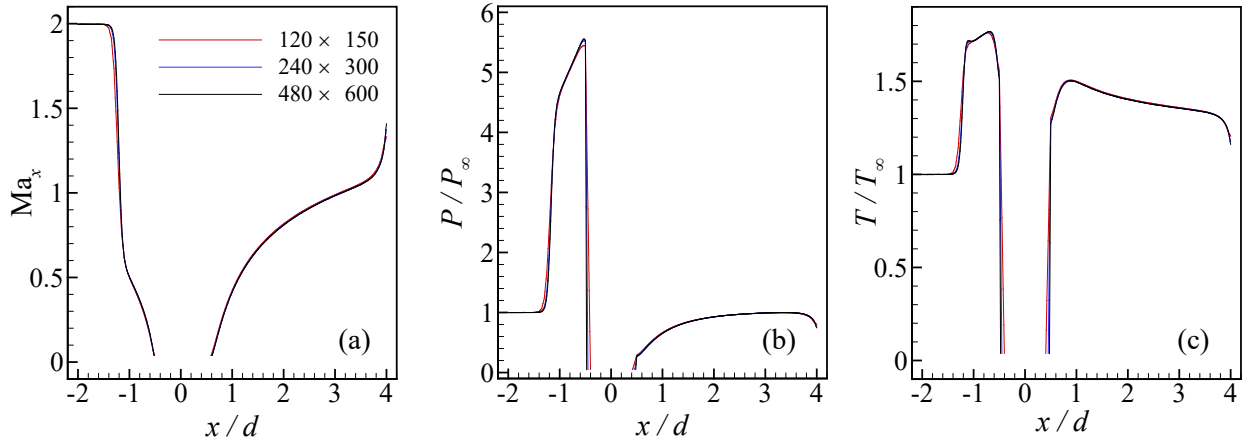

Figure 15. Variations in dimensionless quantities along the symmetry line for the supersonic rarefied gas flow around a cylinder: (a) Mach number, (b) pressure, and (c) temperature.

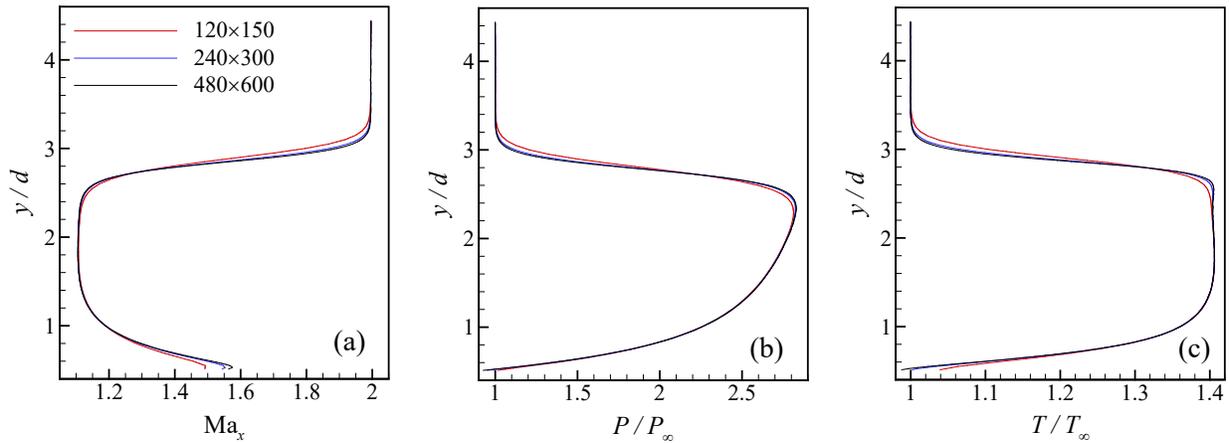

Figure 16. Variation in dimensionless quantities across the vertical section $x = 0$ in the supersonic flow around the cylinder: (a) Mach number, (b) pressure, and (c) temperature.

Figure 17 and Figure 18 show the relative differences in predicted pressure and temperature values obtained from a coarse grid (120×150) and a medium grid (240×300) in comparison to values from simulation using the 480×600 grid as reference. Results from the 120×150 grid show significant deviations across a considerable portion of the simulation domain, indicating its inadequacy in those regions. However, in the free flow areas (upstream of the shock wave), the 120×150 grid provides satisfactory accuracy. Hence, larger cell sizes similar to those in the 120×150 grid can be used in the free flow region. Differences between the results obtained from simulations using the 240×300 and 480×600 grids are evident, with notable differences observed around the shock wave and near the surface of the cylinder. To ensure an adequate level of accuracy in these regions, the use of grid cells as small as those in the 480×600 grid is required. Conversely, in other regions, cells matching the resolutions of the 120×150 or 240×300 grids appear suitable. Since determining the optimal grid is not the primary focus of this study, 240×300 grid cells are employed in those regions to maintain simulation accuracy.

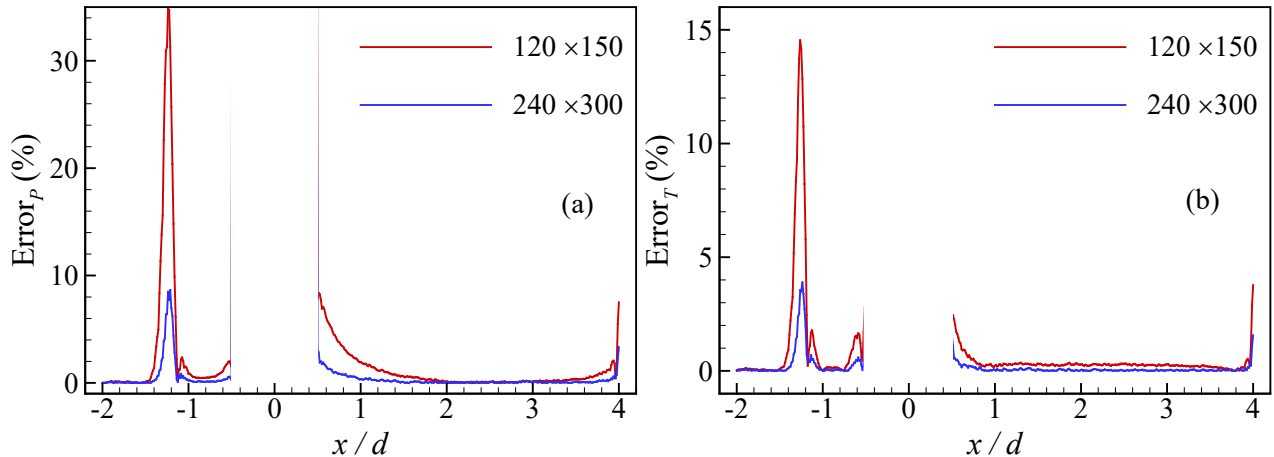

Figure 17. Distribution of the relative errors in predicting dimensionless quantities along the symmetry line in the supersonic flow around the cylinder: (a) pressure error, and (b) temperature error.

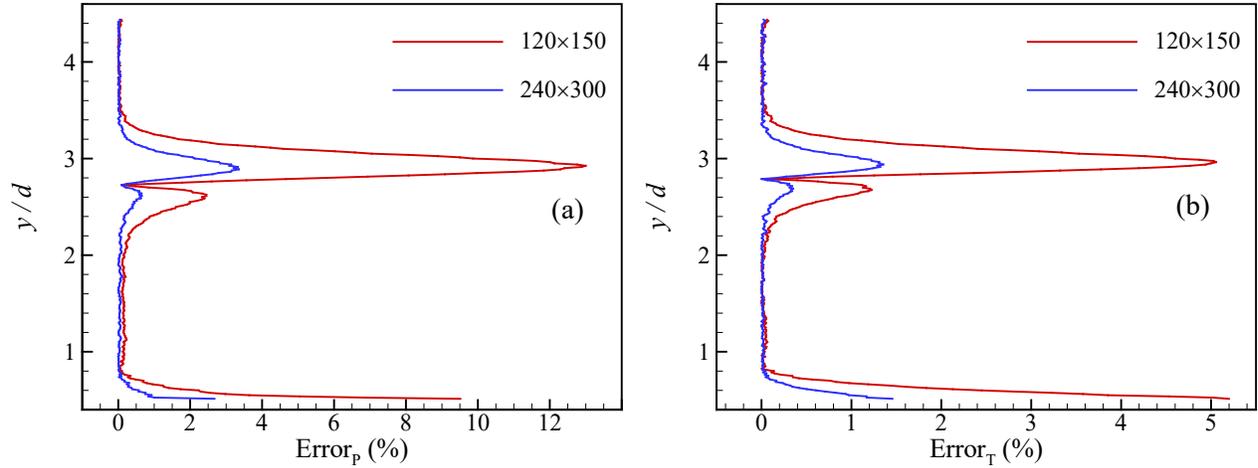

Figure 18. Distribution of the relative errors in predicting dimensionless quantities across the vertical section $x = 0$ in the supersonic flow around the cylinder: (a) pressure error, and (b) temperature error.

A non-uniform grid was employed in the simulations to assess the effectiveness of the FPPC technique. The non-uniform grid was generated based on a uniform 120×150 grid. Using the results obtained from the simulation with a uniform 480×600 grid, the position of the shock was approximated. Cells located in the downstream of the shock wave as well as those located in upstream within a distance of $0.325d$ from the shock wave were subdivided into four cells based on the octree grid refinement approach. Another level of grid refinement was applied to cells located adjacent to the shock wave, within distances of $0.125d$ (in downstream) and $0.225d$ (in upstream) from it. Additionally, this refinement was extended to cells within a distance of $d / 2$ from the surface of the cylinder. This refinement strategy was applied through a comparative analysis of results obtained using different uniform grids. Figure 19 shows the resulting computational grid achieved after performing the local grid refinement.

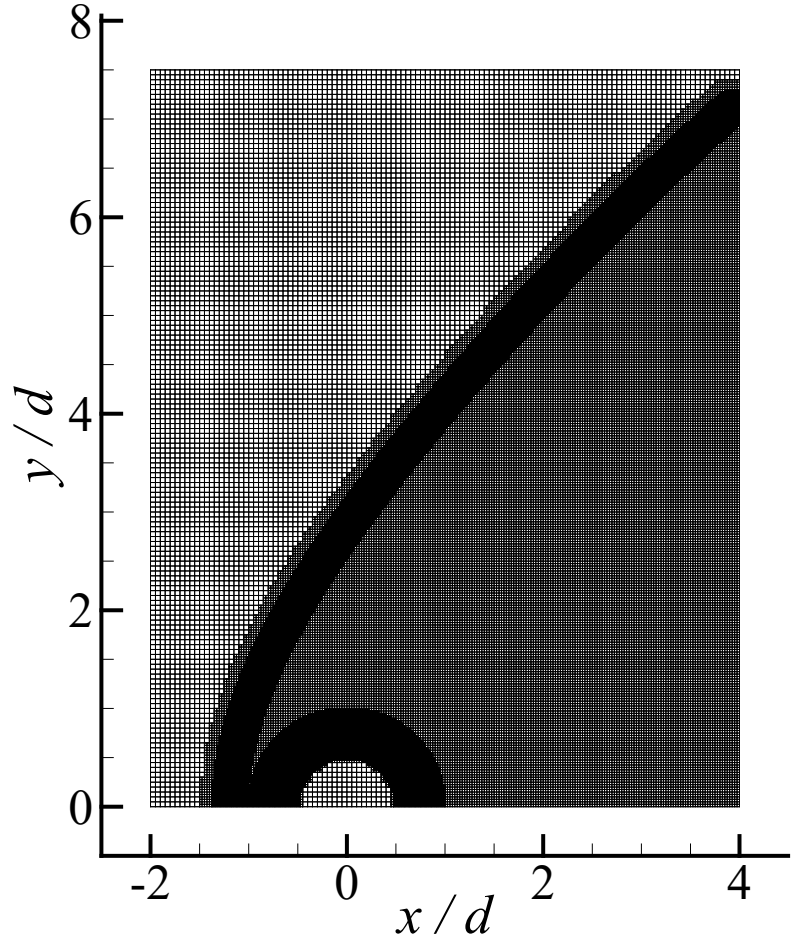

Figure 19. The computational grid generated using local grid refinement to study the supersonic rarefied airflow over a cylinder.

A simulation was conducted using a locally refined grid without applying the FPPC technique. The value of $F_{num}$ in this simulation, with the finest cells comparable to a uniform 480×600 grid, was chosen accordingly. The resulting values for $c_{x,HM}$ and $c_{y,HM}$ were 1.486 and 0.539, respectively. Comparing these results with the data in Table 3 shows negligible differences between this simulation and the one using a uniform 480×600 grid, demonstrating that the proposed grid refinement is adequate for achieving the desired accuracy. Subsequently, the FPPC technique was applied to the locally refined grid, ensuring approximately twenty particles per cell. In this case, the predicted values for $c_{x,HM}$ and $c_{y,HM}$ are 1.487 and 0.539, respectively. Negligible differences (less than 1%) were observed between the results obtained using the conventional DSMC implementation and those using the enhanced implementation with the FPPC technique. Furthermore, contours of Mach number (Ma), pressure ($P^* = P/P_\infty$), and temperature ($T^* = T/T_\infty$) obtained from simulations with FPPC (upper part) and without FPPC (lower part), shown in Figure 20, demonstrate a reasonable agreement between the two implementations. Moreover, the relative differences between the results obtained from simulations using the FPPC technique and those using the conventional DSMC method were calculated using Eq. 7, and the results are shown in Figure 21. The findings show that the greatest relative differences occur in the shock region. Specifically, the maximum difference in predicting pressure in this region is consistently less than 4% in the horizontal direction and less than 2.5% in the vertical direction. The differences in predicting temperature are even smaller, with a maximum difference of less than 2% in the horizontal direction and around 1% in the vertical direction. The higher maximum differences in the horizontal direction compared to the vertical direction can be attributed to the stronger shock in the front region. The bow shock structure causes the highest shock intensity at the front of the body, where the shock is vertical. Moving away from this region, the shock becomes oblique, and its intensity decreases. Near the surface of the cylinder, the differences are generally less than 0.5%. Therefore, it can be concluded that employing the FPPC technique for external rarefied flow problems also yields results with the desired accuracy.

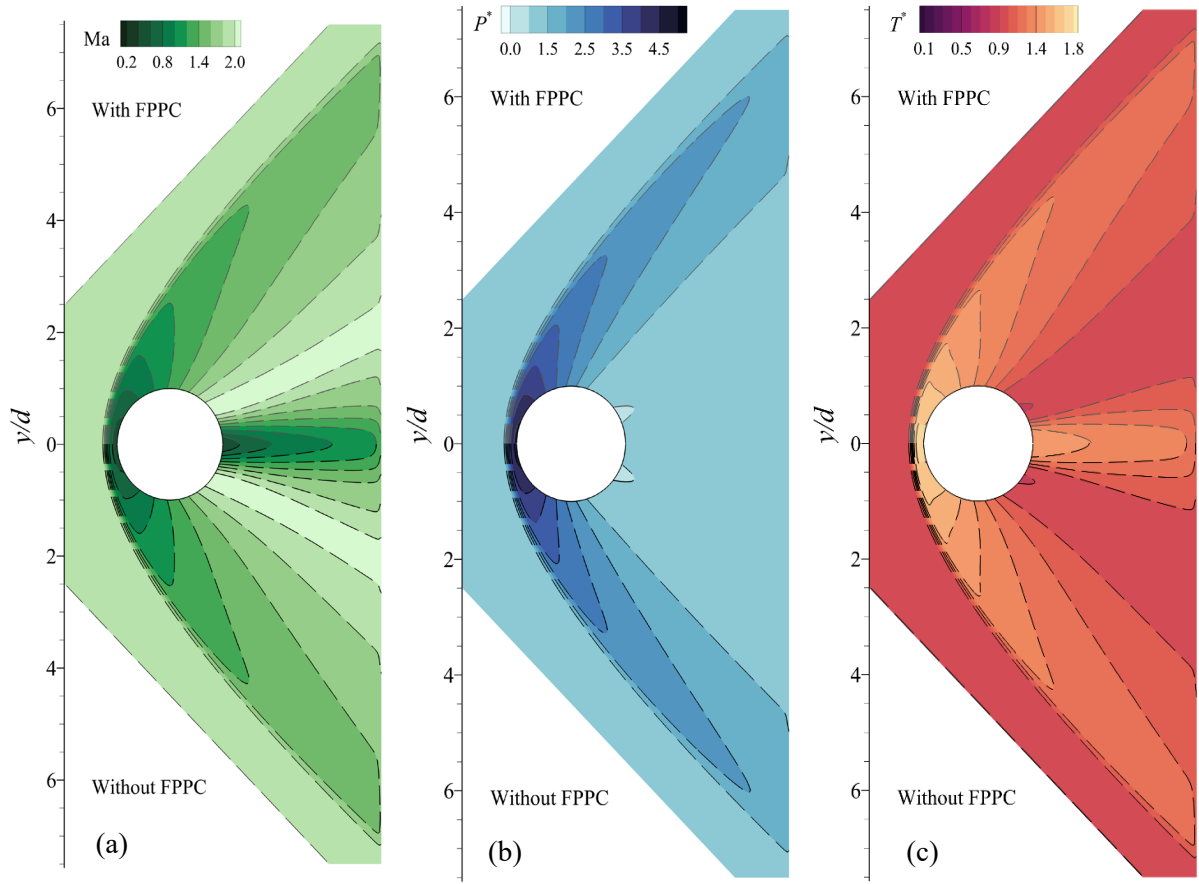

Figure 20. Contours of (a) Mach number, (b) pressure ($P^* = P / P_\infty$), and (c) temperature ($T^* = T / T_\infty$) obtained from DSMC simulations of supersonic airflow over a cylinder with FPPC (upper part) and without FPPC (lower part).

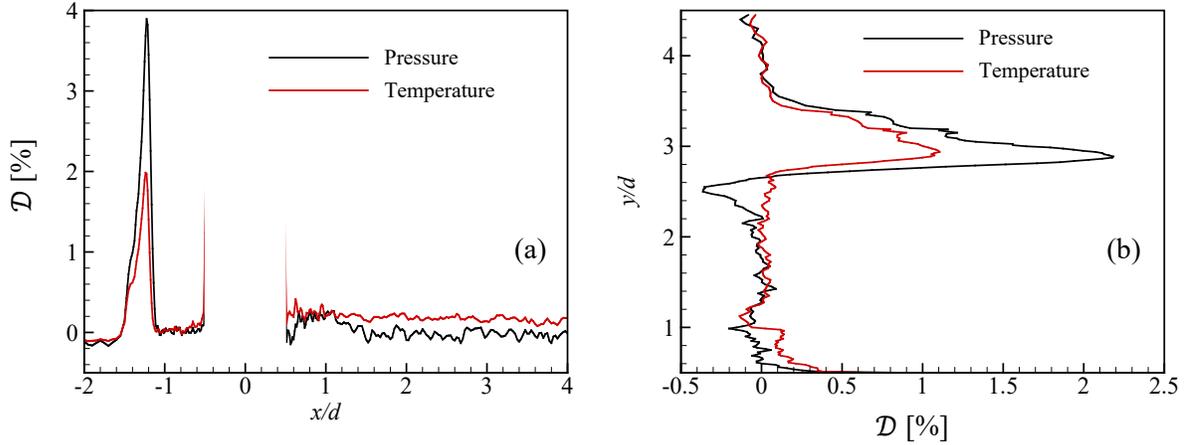

Figure 21. Relative difference between the results obtained from simulations using the FPPC technique and those using the conventional DSMC method for pressure and temperature: (a) $y/d = 0$, and (b) $x/d = 0$.

Table 4 presents parameters related to the computational efficiency of three simulations: uniform 480×600 grid, locally refined grid without the FPPC technique, and locally refined grid with the FPPC technique. A complete simulation of $3\times10^6$ time-steps (encompassing both initial and final stages) required 179.8 hours for the uniform grid without FPPC, 145.3 hours for the locally refined grid without FPPC, and 57.5 hours for the locally refined grid with FPPC. However, for a more accurate comparison, it is advisable to consider the CPU time per time-step during the final stage, which includes both particle dynamics and the sampling process. In this regard, employing local grid refinement along with the FPPC technique resulted in a 3.73-fold reduction in the number of particles and a 2.69-fold reduction in CPU time compared to using the locally refined grid without the FPPC technique. Additionally, compared to the uniform 480×600 grid, there were reductions of approximately 3.73 times in the number of particles and 3.35 times in CPU time.

Table 4. The effects of three different approaches, including a uniform fine grid, a locally refined grid without the FPPC idea, and a locally refined grid with the FPPC idea on the computational cost in the problem of supersonic flow over a cylinder

| Case | Number of Particles | CPU time for a complete simulation (h) | CPU time per time step (ms) | CPU time per time step per particle (μs) |
|---|---|---|---|---|
| Uniform grid 480×600 without FPPC | 4935000 | 179.8 | 218 | 44.2 |
| Locally refined grid without FPPC | 4933000 | 145.3 | 175 | 35.5 |
| Locally refined grid with FPPC | 1323000 | 57.5 | 65 | 49 |

## 5. Conclusions

The present study aimed to enhance the efficiency of direct simulation Monte Carlo (DSMC) simulations for studying vacuum gas dynamics by employing the fixed particle per cell (FPPC) technique in the framework of SPARTA solver. The significance of this work lies in its contribution to advancing the computational efficiency of rarefied gas flow simulations. By employing the FPPC technique, which enforces a fixed number of simulator particles across all computational cells, the present study addresses challenges associated with conventional DSMC implementations, particularly in regions with strong local gradients. Unlike prior "particle count control" methodologies employed in existing literature, the method proposed in the present work eliminates the necessity for real-time adjustment of particle weights. Employing an initial simulation with a reduced number of particles offers dual advantages: facilitating the estimation of weight factors and establishing an appropriate initial condition for the main simulation through a particle scaling procedure. The latter reduces the time required to reach steady-state conditions before starting the sampling procedure.

The findings indicate that the enhanced implementation of the DSMC method with the FPPC technique yields results comparable to conventional methods while significantly reducing computational costs. The performance of the proposed approach in simulating vacuum gas dynamics in microchannels, micromixers and supersonic flow over a cylinder, was investigated. Firstly, validation against conventional DSMC methods showed close agreement, affirming the reliability of the enhanced SPARTA solver. Secondly, for micromixer flow simulations, local grid refinement coupled with the FPPC technique enabled accurate prediction of high-gradient field quantities while reducing computational costs. Thirdly, in supersonic flow over a cylinder, the application of local grid refinement coupled with the FPCC technique resulted in preserving accuracy near shock waves and the cylinder surface at a reduced computational cost. The results show the importance of local grid refinement in regions with significant gradients, facilitated by the FPPC technique, to maintain simulation accuracy. Moreover, the study highlights the potential of non-uniform grids with local refinement to achieve comparable results to uniform grids while reducing computational overhead.

The implementation of the FPPC technique demonstrated a significant reduction in computational resources while maintaining accuracy, making it a viable approach for rarefied gas flow simulations. Specifically, employing the FPPC technique led to a substantial decrease in the total number of simulator particles in DSMC simulations and CPU time compared to conventional DSMC implementations.

**CRediT authorship contribution statement**

**Moslem Sabouri**: Conceptualization, Methodology, Software, Validation, Formal analysis, Investigation, Resources, Data Curation, Writing—Original Draft, Writing—Review & Editing, Visualization, Project administration. **Ramin Zakeri**: Conceptualization, Methodology, Software, Validation, Formal analysis, Investigation, Resources, Data Curation, Writing—Original Draft, Writing—Review & Editing, Visualization, Project administration. **Amin Ebrahimi**: Conceptualization, Methodology, Validation, Formal analysis, Investigation, Resources, Writing—Original Draft, Writing—Review & Editing, Visualization.

**Declaration of competing interest**

The authors state that they do not have any conflicting interests associated with this publication.

**Data availability**

The paper includes representative samples of the research data. Additional datasets produced during the study can be obtained from the authors upon reasonable request.

## Nomenclature

### Symbols

| | | | |
|---|---|---|---|
| $c$ | Dimensionless force coefficient [-] | $P_r$ | Inlet-to-outlet pressure ratio [-] |
| $d$ | Diameter [-] | $PPC_i$ | Initial value of the number of particles per cell in cell $i$ [-] |
| $D_{AB}$ | Binary diffusion coefficient [m$^2$ s$^{-1}$] | $PPC_t$ | Target value for the number of particles per cell [-] |
| $\boldsymbol{F}$ | Force [N] | $s$ | Scaling parameter [-] |
| $F_{num}$ | Equivalent number of molecules [-] | $T$ | Temperature [K] |
| $H$ | Height of the channel [m] | $u$ | Velocity magnitude [m s$^{-1}$] |
| $\boldsymbol{J}_A$ | Diffusive mass flux [kg m$^{-2}$ s$^{-1}$] | $V_{cell}$ | Cell volume [m$^3$] |
| $\boldsymbol{J}_A^*$ | Dimensionless diffusive mass flux [-] | $w$ | Weight parameter [-] |
| Kn | Knudsen number [-] | $x, y$ | Cartesian coordinates [m] |
| $L$ | Length of the channel [m] | $\mu$ | Dynamic viscosity [Pa.s] |
| Ma | Mach number [-] | $\rho$ | Density [kg m$^{-3}$] |
| $n$ | Number density (number of real molecules per unit volume) [m$^{-3}$] | $\sigma$ | Accommodation coefficient [-] |
| $P$ | Pressure [Pa] | $\lambda$ | Molecular mean free path [m] |
| $\hat{P}$ | Non-dimensional pressure distribution [-] | | |

### Subscripts

| | | | |
|---|---|---|---|
| e | Outlet | $i$ | Cell number |
| HM | Half model | lin | Linear variation |
| in | Inlet | ∞ | Free stream condition |

### Acronyms

| | | | |
|---|---|---|---|
| CFD | Computational fluid dynamics | LD | Low diffusion |
| CPU | Central processing unit | PIC | Particle-in-cell |
| DSMC | Direct simulation Monte Carlo | PPC | Number of particles per cell |
| EPSM | Equilibrium particle simulation method | SPARTA | Stochastic parallel rarefied-gas time-accurate analyser |
| FPPC | Fixed particle per cell | VSS | Variable soft sphere (molecular collision model) |
| GPL | General public licenses | | |